\documentclass[10pt]{iopart}

\usepackage{graphicx}
\usepackage{dcolumn}
\usepackage{multirow}
\usepackage{bm}
\usepackage{amsfonts,amssymb}
\usepackage{xcolor}
\newcommand\Ru{{\rm Re}}
\newcommand\Rm{{\rm Rm}}
\begin{document}
\title[Solar Wind in Shell Model]{Transformation of solar wind energy and helicity spectra in the frame of magnetohydrodynamics shell modeling}

\author{I. Dukanov$^{1,3}$, E. Yushkov$^{1}$,
P. Frick$^2$ and D. Sokoloff$^{1,3}$}
\address{$^1$ Physics Department of Lomonosov Moscow State University, Moscow, Russia}
\address{$^2$ Institute of Continuous Media Mechanics UB RAS, Perm, Russia}
\address{$^3$ Pushkov Institute of Terrestrial Magnetism, Ionosphere and Radio Wave Propagation RAS, Troitsk, Russia}
\ead{idukanov@icloud.com}

\vspace{10pt}
%\begin{indented}
%\item[]September 2025
%\end{indented}

\begin{abstract}
Based on the data recorded during the Parker Solar Probe mission, it can be suggested that there is no balance between kinetic and magnetic energy in the vicinity of the Sun. The spectra collected at different radial distances show an energy transfer from one component to another, followed by a change in inertial range slope and large-scale break shifted towards lower wavenumbers. Using the shell (cascade) modeling approach, we attempt to explain and understand this observed spectra transformation, considering a free-decay turbulence process and modeling a simple transition to a quasi-stationary equilibrium between the magnetic and kinetic energies. Varying the parameters of turbulence mirror asymmetry (helicity), we numerically simulate changes in spectral indices, large-scale shift position, and inertial range transformations, comparing our model results with spacecraft observations of solar wind turbulence. We present and discuss the chaotic nature of the spectral magnetic helicity distributions, as well as the fact that helicity can accumulate at large scales. The applicability of a simple MHD shell model to repeat the key features of solar wind turbulence dynamics offers wide possibilities for its further use.
\end{abstract}

\vspace{2pc}

%\submitto{\PR}
%\vspace{2pc}
%\noindent{\it Accepted }: solar wind, energy spectra, free decay, MHD shell model, helicity dynamics

\maketitle
\ioptwocol

%%%%%%%%%%%%%%%%%%%%%%%%%%%%%%%%%%%%%%%%%%%%%%%%%%%%%%%%%
\section{Introduction}
The free decay of turbulence is a classical problem of the hydrodynamics theory. In the absence of an external forcing, when the turbulent energy transport across the spectrum is determined only by its initial distribution and small-scale dissipation, a free decay demonstrates the familiar pattern of Kolmogorov's inertial range formation. Such evolution is quite typical, if there is a sufficient initial energy reserve, weak dissipation, and preferably near-classical local homogeneity and isotropy assumptions. Within the context of magnetohydrodynamics (MHD), the free decay problem becomes more complex, as it depends not only on the ratio between magnetic and kinetic energies, but also on integral characteristics of small-scale turbulence, such as cross helicity (a measure of fields correlation) and magnetic helicity (a measure of mirror asymmetry). The question as to when a familiar spectral pattern with an inertial power range forms in the MHD cascade, and when it does not, has not been fully explored yet, and this study aims to address it. 

Conducting laboratory experiments with the MHD free-decay cascade is an extremely challenging task due to boundary issues, isotropy conditions  and achieving sufficiently high Reynolds numbers. An interesting and potentially unique example, analogous to the classic problem of ``water pounding in a mortar'', demonstrates that the laboratory results can significantly vary depending on whether the grid is installed first and then the field is turned on, or {\it vice versa} \cite{voronchikhin1985experimental}. However, nature has provided us with a convenient testing ground for observing decaying MHD turbulence with very large values of governing parameters -- the interplanetary medium filled with solar wind (SW) plasma. Since the first measurements of interplanetary plasma by the Soviet automatic station Luna-1 in 1959, the properties of SW turbulence have been quite continuously monitored and studied, see, e.g., \cite{harvey2007soviet}. Moreover, with the launch of the Parker Solar Probe spacecraft mission in 2018 \cite{raouafi2023parker}, it has become possible to observe the turbulence decay evolution starting from the closest to the Sun distances, near $R=0.04$ astronomical units (au). The main limitation of this ``testing ground'' is that we cannot manipulate the input parameters, such as the efficiency of dissipation, initial energy distributions and asymmetry - we can only observe what happens.

A good compromise would be to develop a digital model that replicates the main features of turbulence dynamics in the solar wind. This model could then be used to identify the key features of these dynamics, vary input parameters over a wide range, and analyze the influence of difficult-to-measure cross and magnetic helicity distributions. \textcolor{black}{However, direct numerical simulations (DNS) of the evolution of fully developed SW turbulence are very difficult due to the large range of wavenumbers which should be resolved. For example, three orders of magnitude in the SW spectral range (see, e.g., \cite{chen2020evolution}) and two additional orders of wavenumbers for the scales before and after the inertial range lead to at least $10^{15}$ grid points in the spatial numerical grid.} So, instead of DNS methods, other modeling techniques and simplifications should be used.

One of the approaches proposed in the middle of the last century for modeling the fully developed turbulence (including the MHD turbulence) was the so-called  shell method, in which the MHD equations were represented in $k$-space by a finite number of spectral shells. The main idea is that the Fourier images of the velocity and magnetic field vectors can be replaced by collective scalar variables, whose squares are associated with kinetic/magnetic energy densities within these shells. In the MHD equations describing energy transport from shell to shell, the Fourier transformations of nonlinear terms are assimilated by quadratic nonlinearities that describe shell-to-shell local interactions, satisfying 3D MHD conservation laws in the nondissipation limit. These models, whose formalism was introduced and formulated in the 1970s by A.~Obukhov \cite{Obukhov1971} and E.~Lorenz \cite{Lorenz1972}, are still used to simplify the description of turbulent energy and helicity transport in various turbulent systems. 

In this paper, the shell-based approach, simulating the SW turbulence, is employed not to provide a detailed explanation of how the SW spectra evolve, as there are numerous papers dedicated to this topic, primarily based on direct spacecraft observations, see, e.g., the review \cite{alexandrova2013solar}, but to explore the potential of using a relatively simple and abstract shell model to understand and roughly simulate the evolution of SW turbulence near the Sun. Thus, the SW primarily provides the initial data for the numerical simulations and verifies the results. Besides, this study can be considered as a continuation of previous investigations of SW with the shell approaches, which were described by Bruno and Carbone in \cite{bruno2013solar, bruno2016turbulence}. The main difference between our work and previous ones is that we do not use a model specialized exactly for the SW and do not try to replicate exactly the relationship between magnetic and kinetic energies or the stochastic effects of SW intermittency, see, e.g., \cite{ditlevsen1997cascades}. Considering the evolution of spectra over short time intervals, we do not take into account the external magnetic/kinetic energy/helicity inputs at both large and small scales. This means that the focus of our study is only on the free-decay of turbulence generated in the vicinity of the star. Another difference is that we do not apply the generally used Gledzer-Ohkitani-Yamada (GOY) models, where the helicity sign is connected with the particular shell number, see, e.g., \cite{brandenburg2023turbulence}. Instead, in our approach the sign of helicity can take any value in each shell. So, by varying different initial helicity distributions, we search for the most appropriate dynamics compared to the real situation.

Choosing such a shell model, we aim to describe the dynamics of the main spectra features rather than the exact SW energy balance. In particular, we suggest that the large-scale breaks observed in turbulent spectra are not caused by changing external conditions or forward/reverse energy cascades, but by the natural transfer of magnetic energy to kinetic energy, which affects larger and larger scales over time. Additionally, we demonstrate that the spectral indices ranging from $-3/2$ to $-5/3$ also occur during this natural process of magnetic-kinetic energy transfer. Moreover, if the magnetic spectrum index changes more rapidly and with greater fluctuations, the kinetic spectrum uniformly transforms at distances of several astronomical units (au). Finally, we show that, within the shell modeling, the non-monotonic structure of the helicity spectral density is formed naturally on the whole spectral range, which corresponds to many spacecraft observations \cite{brandenburg2011scale}. The magnetic helicity, being a conserved quantity, tends to condense at largest scales with mean values, which can be far from zero.

It is also worth to note that the simplicity of the shell model prevents us from claiming that the simulated evolution fully matches the observed SW data. Our results show which features of SW evolution are basic and observable, even within the simplest framework of the shell approach. The structure of the paper follows a standard format:  it describes the initial collection of input data, data preparation for shell modeling, numerical simulation results, and comparison of the outcomes with actual observed data.

\section{MHD shell approach}
The shell approach for fully developed turbulence is based on the idea to convert the equations of motion into the spectral space and then reduce them to a set of ODE written for some collective variables, each of which represents fluctuations in a whole spectral sub-interval (shell) $k_n<k<k_{n+1}$. The points $k_n=\lambda^n$ follow a geometric progression, that is, they are evenly distributed on a logarithmic scale. The MHD shell equations are written for magnetic and velocity fields described by  corresponding collective variables:
$U_{n}=a_{n} + ib_{n}$ and $B_{n}=c_{n} + id_{n}$, scalar due to the assumption of isotropy of turbulent fields.   
The squares of these collective variables are associated with the kinetic $E_u$ and magnetic $E_b$ energies in  corresponding shells, their sum describes the total energy $E$:
\begin{equation}\label{Eq01}
E =E_u+E_b= \sum_n(a_n^2+b_n^2)+\sum_n(c_n^2+d_n^2).
\end{equation}
The analogues of the cross-helicity and magnetic helicity are defined by quadratic forms
\begin{equation}\label{Eq01b}
H_b =\sum_n c_{n}d_{n}/k_{n}, \;\;
H_c =\sum_n (a_n c_n + b_nd_n).
\end{equation}
The general dimensionless form of shell equations is
\begin{equation}\label{Eq04}
d_t U_n={W_n}({\bf U},{\bf U})-{W_n}({\bf B},{\bf B})-k_n^2U_n/{\rm Re}, 
\end{equation}
\begin{equation}\label{Eq04b}
d_t B_n={W_n}({\bf U},{\bf B})-{W_n}({\bf B},{\bf U})-k_n^2B_n/{\rm Rm},
\end{equation}
with two parameters ${\Ru}$ and ${\Rm}$ corresponding to Reynolds numbers and determining the kinetic and magnetic  dissipation. The choice of a specific model is determined by the type of nonlinear terms $W_n({\bf X},{\bf Y})$ used (see  \cite{plunian2013shell} for an extended review of different kinds of MHD shell models).

We accept the model introduced in \cite{mizeva2009cross}, in which the nonlinear terms $W_n({\bf X},{\bf Y})$ are defined by a bilinear complex form
\begin{eqnarray*}
W_n({\bf X},{\bf Y})=ik_n[(X_{n-1}Y_{n-1}+X_{n-1}^*Y_{n-1}^*)-\lambda
X_n^*Y_{n+1}^*\nonumber\\
-\frac{\lambda^2}{2}(X_n Y_{n+1}+X_{n+1}Y_n+X_nY_{n+1}^*+X^*_{n+1}Y_n)\nonumber\\
-\frac{\lambda}{2}(X_{n-1}^*Y_{n-1}-X_{n-1}Y_{n-1}^*)
+\lambda X_n^*Y_{n+1}] \nonumber
\\ -i k_n \lambda^{-5/2}[\frac{1}{2}(X_{n-1}Y_n+X_nY_{n-1})+\lambda
X^*_nY_{n-1}^*\nonumber\\-\lambda^2(X_{n+1}Y_{n+1}+X_{n+1}^*Y_{n+1}^*)+\frac{1}{2}(X_nY_{n-1}^*+X_{n-1}^*Y_{n})
\nonumber\\
-\lambda X^*_nY_{n-1}+\frac{\lambda}{2}(X_{n+1}^*Y_{n+1}-X_{n+1}Y_{n+1}^*)],\phantom{XX}
\end{eqnarray*}
with  $\lambda=1.618$ and $-10\leq n\leq 30$ ($41$ spectral shells).

We would like to highlight one specific feature of the shell model that is relevant to the results obtained. The conservation laws that define the type of nonlinear terms are expressed in terms of (\ref{Eq01})-(\ref{Eq01b}). The definitions of energy and cross-helicity are obvious and are used in all shell models, while the magnetic helicity can be introduced in different ways.  It is worth noting that helicities are pseudoscalars and the sign of them depends on the definition of the coordinate system. In the shell model the expressions for helicity can be chosen with either a positive or negative sign. The shell model remains unchanged if the sign is reversed, so the helicity demonstrated below can also be considered with an opposite sign. This is significant, for instance, because we show below that the value $kH_b/E_b$ can have a different sign compared to spacecraft experiment investigations \cite{brandenburg2011scale}.

\section{Input data for shell modeling extracted from solar wind spacecraft observations}
Intending to simulate the evolution of the solar wind even in the frame of a very reduced model, which ignores the details of the spatial distributions of all the characteristics, one needs some amount of input information, which includes the estimates of typical hydrodynamic and magnetic Reynolds numbers, $\Rm$ and $\Ru$, the characteristics of the mirror asymmetry of turbulence (helicity $H_u$) and magnetic field (cross-helicity $H_c$ and magnetic helicity $H_b$) associated with the conservation laws of ideal MHD, and the initial spectral densities of magnetic and kinetic energies.

%%%%%%%%%%%%%%%%%%%%%%%%%%%%%%%%%%%%%%%%%%%%%%%
\begin{figure}[t]
\centering
\includegraphics[width = 0.49 \textwidth]{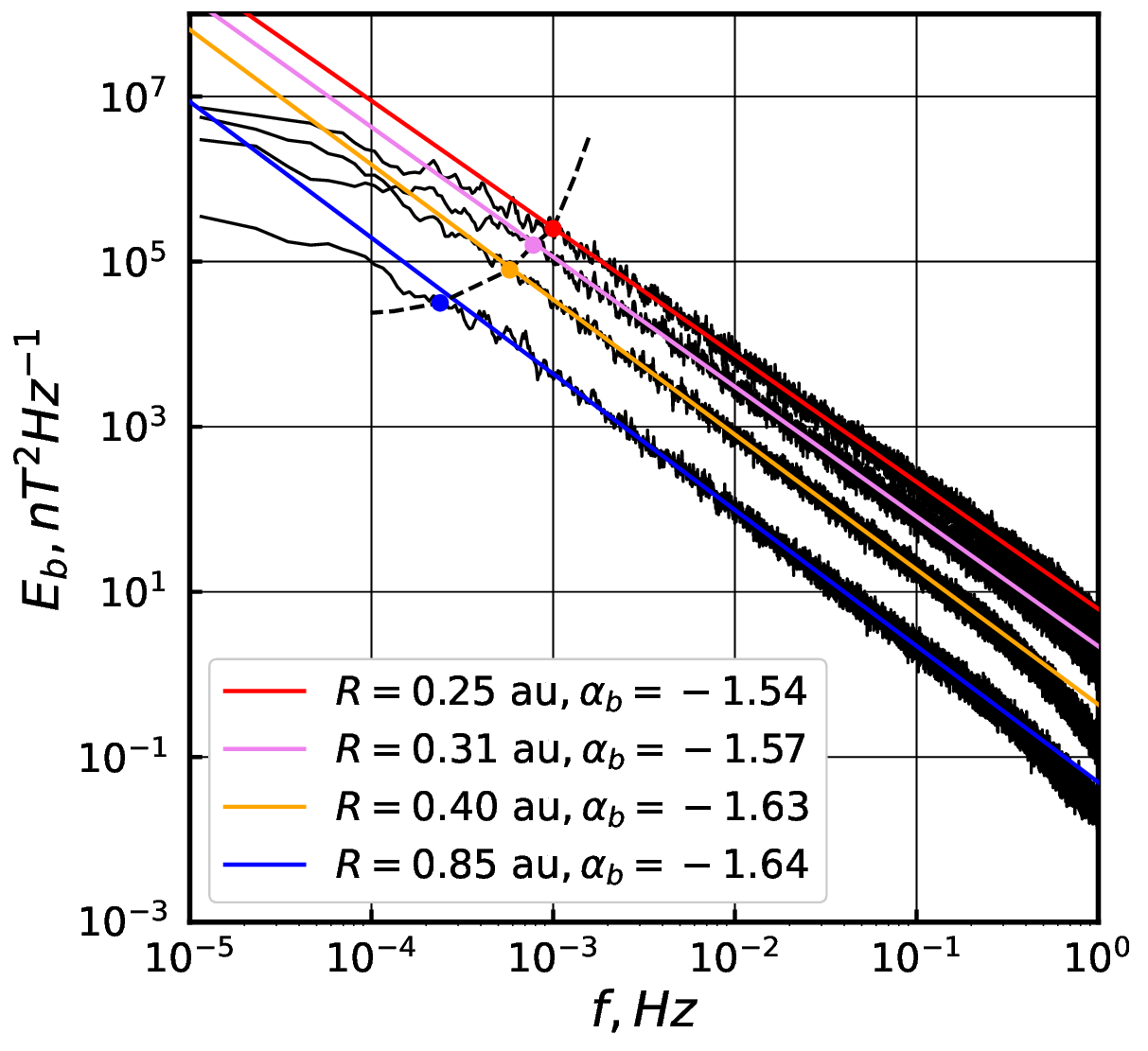}
\includegraphics[width = 0.49 \textwidth]{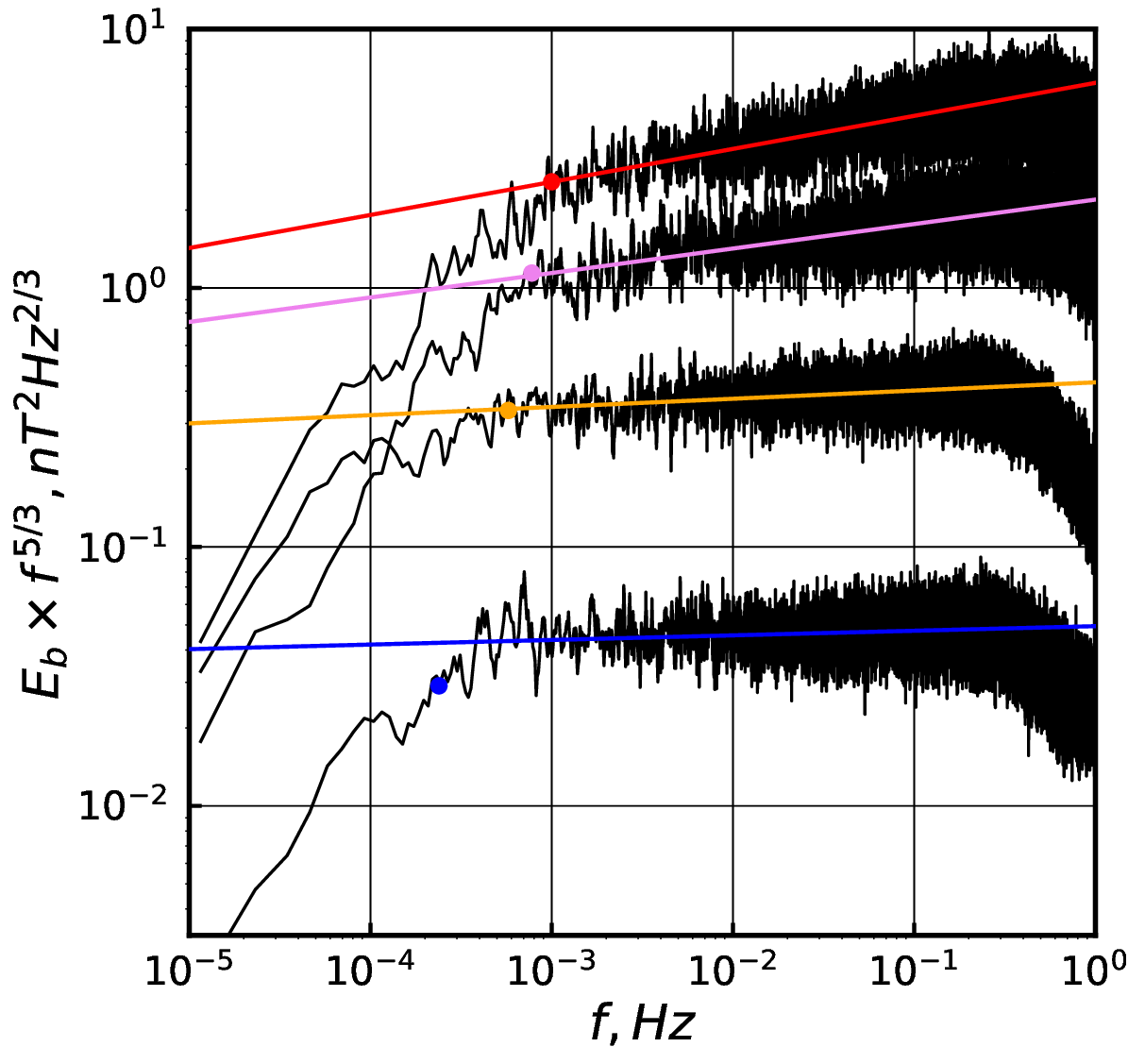}
\caption{Magnetic field energy spectra (top) and compensated energy spectra (bottom) obtained by averaging five daily periodograms at various heliocentric distances: $R=0.25$ (red), $0.31$ (pink), $0.40$ (orange) and $0.85$ (blue) au. The spectra demonstrate the markers of solar wind turbulence evolution discussed later: energy decrease, change of spectral index and leftward shift of the large-scale break marked by colored circles.}
\label{Fig01}
\end{figure}
%%%%%%%%%%%%%%%%%%%%%%%%%%%%%%%%%%%%%%%%%%%%%%%

Of these, only the energy densities that are reconstructed near the Sun from the PSP data (NASA's CDAWeb resources) can be considered reliable. The energy distributions are reconstructed from the time series of magnetic field components and ion bulk velocity (FIELDS and SWEAP instruments correspondingly, see \cite{Bale_MacDowal_Koval_Pulupa_Quinn_Schroeder_2020,Kasper_Stevens_Case_Korreck_2020}). The magnetometer data in RTN coordinates are collected with the resolution of 218.453 ms (as PSP FIELDS 4 samples per cycle cadence). Velocity data do not have a constant resolution (proton bulk velocity derived from 1D Maxwellian fitting), so they are manually processed to fill gaps in the measurements with averaged values (the arithmetic mean of the two preceding values is inserted where data are missing). The measurements used in this study are selected to be free from the signs of solar flares and shock waves, and have mean field variations that are smaller than their fluctuations. Weekly time intervals of $20$-$27.01$, $18$-$25.01$, $14$-$21.01$ for 2020 and $09$-$16.12$ for 2019 were chosen during the PSP's journey toward the Sun. The selected intervals correspond to the heliocentric distances of $R=0.25$, $0.31$, $0.40$ and $0.85$ au obtained from HELIO1DAY POSITION sensor measurements, see \cite{Papitashvili_2023}, averaged over each weekly interval. We limited ourselves to four time cuts because they quite clearly demonstrate the typical evolution of energy spectra, which we try to reproduce numerically.

The examples of magnetic energy densities recovered from PSP data for selected time intervals are shown in Figure~\ref{Fig01}. These examples clearly demonstrate a decrease in total energy with increasing distance from the Sun associated with dissipation and expansion of the solar wind plasma. To show the weak but monotonous change of the spectral slope in the inertial range (the spectral index $\alpha_b$ is indicated in the legend), we present on the bottom panel the spectra compensated by the Kolmogorov ``-5/3'' power law. The large-scale break of the inertial range is marked in each spectrum by the markers of the corresponding color, which show that the break is shifting to larger scales. This is a fairly typical evolution scenario, which is presented in more detail, for example, in \cite{chen2020evolution} or \cite{mcintyre2023properties}, where, among other things, it is demonstrated that $\alpha_b$ can vary from $-3/2$ to $-5/3$ when passing approximately one au, \textcolor{black}{ and the position of the break point depends on the distance from the Sun as $R^{-1.12 \pm 0.16}$ and on the travel time from the Sun as $t^{-1.25 \pm 0.11}$. A similar power-law behavior of a large-scale break is described in \cite{lotz2023radial}, but with a slightly higher degree $R^{-1.18 \pm 0.02}$.} Note that SW spectra show not only the large-scale but also the small-scale break (near the scale of ion gyro-radius) as can be seen in the right part of Figure~\ref{Fig01}, however, since our study focuses only on the large-scale MHD effects, we do not consider the small-scale break.

Next, to start the shell modeling of MHD spectra evolution, we have to estimate the dissipation effects and Reynolds numbers, $\Ru=VL/\nu$ and $\Rm=VL/\mu$. These numbers characterize the ratio between the product of typical velocity $V$ and scale $L$ to the kinematic viscosity $\nu$ or magnetic viscosity $\mu$, see, e.g.,
\cite{davidson2017introduction}. Unfortunately, currently there are no reliable methods for estimating SW viscosities, therefore, the formula $\Ru \sim (\lambda_C/\lambda_T)^2$ is usually used to estimate Reynolds numbers as the ratio of correlation scale $\lambda_C$ to Taylor scale $\lambda_T$. Both the scales are calculated from the correlation function of the fluctuating magnetic field, but some questions concerning $\lambda_T$ estimation still remain. Moreover, there is no clear rule on how to choose the magnetic Prandtl number, the ratio of $\Ru$ and $\Rm$, see \cite{perez2004empirical, hand2026empirical}. Referring to one of the most comprehensive recent reviews, \cite{wrench2024reynolds},  it can be stated that  the estimates of the magnetic Reynolds number vary over a very wide range: from $10^3-10^4$, e.g., \cite{bandyopadhyay2020direct}, to $10^6-10^8$, e.g., \cite{weygand2009anisotropy}. The complexity of the estimation is related to the fact that the accuracy of the Taylor scale reconstruction depends on the resolution of the spacecraft device. The shorter the time resolution (or spatial one), the more accurately the correlation function can be measured. This is one of the reasons why Taylor scale is best measured for magnetic fields; this is exactly how a magnetometer works. In most investigations, e.g., where the kinetic Reynolds number is estimated, it is actually estimated using the correlation function of the magnetic field, rather than the velocity one \cite{cuesta2022intermittency}, this is why it is usually called the ``effective" Reynolds number (the estimate based on some of the most recent PSP data gives, in particular, $\Ru=10^4$, see \cite{phillips2022taylor}, or $\Ru=1.5\cdot 10^6$ and $\Rm=0.9\cdot 10^6$ from Spektr-R data, see \cite{hand2026empirical}). In our modeling, we tested both Reynolds numbers over the range $10^4-10^7$, but it has turned out that these choices primarily determine the shape of the spectra rather than the overall energy behavior. And since at ${\Rm}<10^{6}$ the inertial range of sufficient extent does not arise, both the kinetic and magnetic Reynolds numbers were set to $10^{6}$. 

%%%%%%%%%%%%%%%%%%%%%%%%%%%%%%%%%%%%%%%%%%%%%%%
\begin{figure}[t]
\centering
\includegraphics[width = 0.48\textwidth]{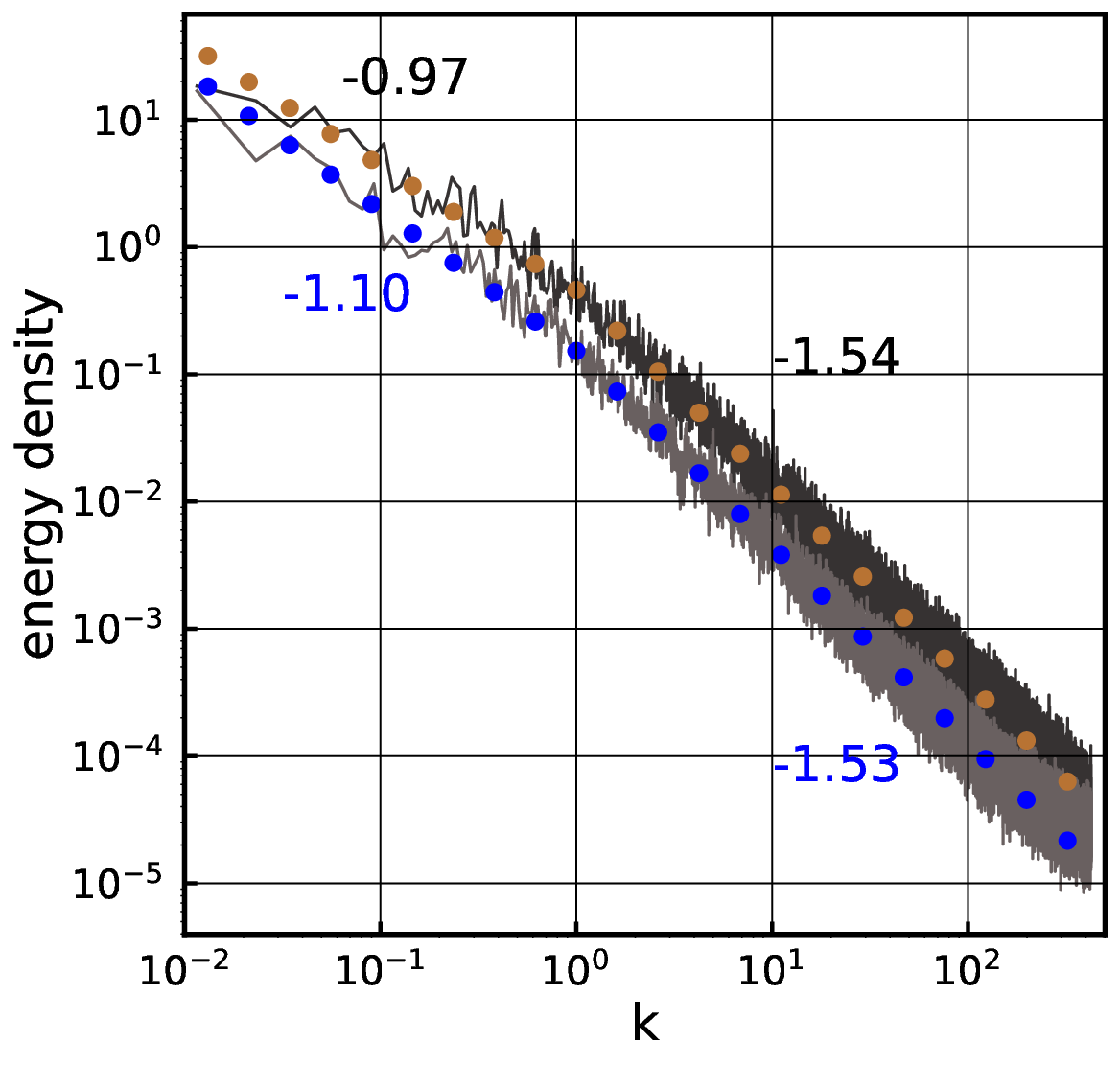}
\includegraphics[width = 0.48\textwidth]{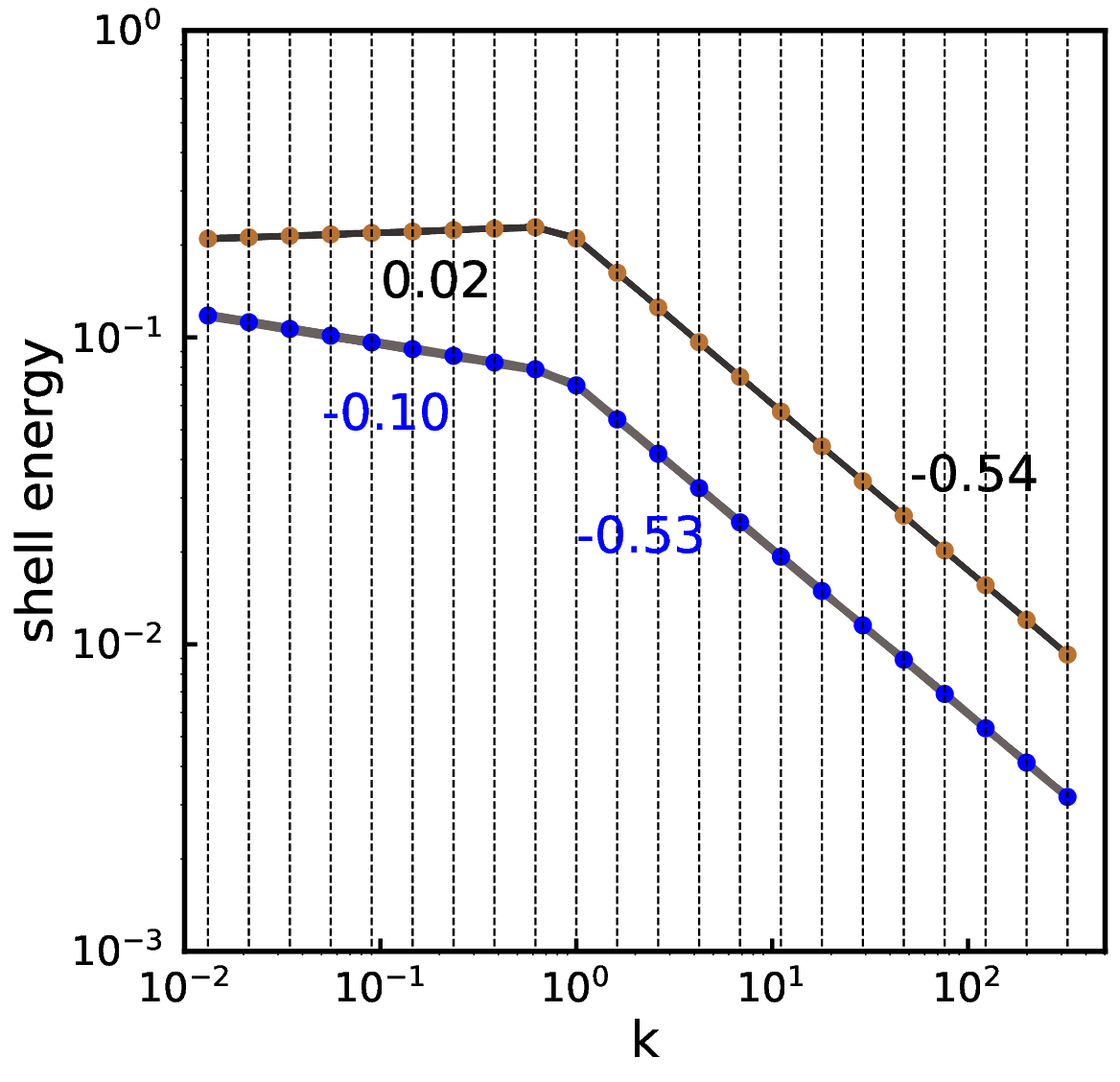}
\centering
\caption{Dimensionless spectra of kinetic (gray) and magnetic (black) energy near the Sun at $R = 0.25$ au are used as the input model data. The colored points show their approximations with two lines. The lower panel displays similar approximating densities integrated for each of the $N$ spectral shells marked by vertical dashed lines and multiple input realizations with one-percent noise \textcolor{black}{(500 realizations for each curve are blurred into thick black and gray lines for magnetic and kinetic energies correspondingly)}.}
\label{Fig02}
\end{figure}
%%%%%%%%%%%%%%%%%%%%%%%%%%%%%%%%%%%%%%%%%%%%%%%

Finally, we have to explain how we made the most questionable estimates, which concern the helicities of SW turbulence. As mentioned earlier, the shell modeling is based on the conservation laws for the dissipationless limit. In 3D MHD, the conservation laws, besides the total energy, include the magnetic and cross helicities. This means that we have to control the level of helicities and, therefore, the mirror asymmetry of turbulence generating the input data for the shell model. There are several studies that examine the cross and magnetic helicities in SW, e.g., \cite{brandenburg2011scale,podesta2011magnetic,markovskii2015statistical,bavassano1998cross}, however, their applications can not be done straightforward like energy because helicities can be both positive or negative, therefore, average values can be far away from an individual realization. Moreover, accurate estimations of the spectral density of helicities require long-term multi-satellite measurements, which are difficult to achieve, especially in the vicinity of the Sun. A significant part of the paper is related to various variants of cross ($H_c$), kinetic ($H_u$) and magnetic ($H_b$) helicities assuming their proportionality to corresponding energies. By varying these parameters, we select those that match better the observed evolution of spectra. We will return to this problem preparing the input model data.

\section{Preparation of initial model data}
When modeling stationary turbulent flows, the statistical poverty of shell models is compensated with extremely long time implementations. Since we are interested in the evolution of the spectral composition of the solar wind, that is, we are modeling an unsteady process, the only way to collect statistically reliable data is to simulate a large set of single implementations, each of which starts with its own random set of initial data corresponding, as far as possible, to observational data at the closest distance from the Sun at which the  measurements were performed.

Thus, we have to generate random sets of initial conditions for the real and imaginary parts of the shell variables $a_n$, $b_n$, $c_n$ and $d_n$, which reproduce in some way the spectral properties of the spacecraft data collected at the closest distance  to the Sun, $R=0.25$ au. We describe the procedure of preparing the data for the magnetic field, $B_n=c_n+id_n$, the same procedure was applied for the velocity field, $U_n=a_n+ib_n$.

First, we converted the original units of spectral energy density, ${\rm nT}^2\cdot{\rm Hz}^{-1}$ and ${\rm (m/s)}^2\cdot{\rm Hz}^{-1}$, to a uniform unit for the SI system, ${\rm J}\cdot {\rm Hz}^{-1}$, which allows us to plot the kinetic and magnetic spectra on the same axis (for the velocity field, we use the proton concentration in the SW plasma at the considered distance, $n_0 = 114 \cdot 10^6$ ${\rm m}^{-3}$). Next, we transform the frequency spectra into spatial ones using the Taylor hypothesis, dividing all frequencies and multiplying all energy densities by the average flow velocity calculated from experimental data, $U_{f}=344$ km/s.  Finally, the wavenumbers are divided by $k_0 = 2\cdot10^{-9}$ ${\rm m}^{-1}$, so that $k=1$ on the dimensionless axis corresponds to the large-scale break of the magnetic field spectrum. The position of the break is determined by the method shown in Figure \ref{Fig01}. Approximating the inertial range with a straight line, we select the point where the line deviates from the data. The dimensionless energy density is measured by the fluctuation energy corresponding to the velocity $U_{en} = 48$  ${\rm km/s}$. \textcolor{black}{The length unit in this case is $L=\pi\cdot10^9$ m, and the unit of time is $T=6.5\cdot10^4$ s. These units correspond to the heliocentric distance of about $R=0.15$ au. Of course, it is important to understand that this correspondence is valid only as a zero-order approximation. Spacecraft observations show that the average SW velocity varies with radial distance, see, e.g., \cite{dakeyo2022statistical,halekas2022radial}, so a more accurate return to dimensional quantities requires a more careful approach. However, for the purposes of our study, the given rough correspondence will be sufficient for several reasons. First, we do not attempt to convert the model data back to dimensional quantities considering only the typical evolution of dimensionless markers. Second, within one au, the average velocity varies from 200 km/s near the Sun to 400 km/s on the Earth’s orbit: approximately the same spread in velocity is observed at a fixed radial distance. Particularly, in our case, we have velocity $U_{f}$, which corresponds to the average velocity within several au. Finally, the shell model used is chosen to be simplified specifically to see the typical evolution, and therefore does not include those factors that would need to be taken into account simultaneously with the bulk velocity changing, in particular, the expansion and cooling of the SW plasma.}

\textcolor{black}{The dimensionless magnetic and kinetic energy densities are presented in the top panel of Figure \ref{Fig02} in black and gray respectively. However, as mentioned earlier, the input data to the shell model are not the densities, but the energies in the corresponding spectral shells, $|B_n|^2$ and $|U_n|^2$. This means that the dimensionless input data must be integrated over each subrange $[k_{n},k_{n+1}]$. To prevent unnecessary errors (resulting from the Fourier transform), we interpolate the input spacecraft data by power-law functions (colored dots in the top panel). The resulting energies integrated over the shells are also highlighted by the similar colored dots in the bottom panel ($[k_{n},k_{n+1}]$ subranges are shown by thin vertical lines). It is for this reason that the power-law dependencies indicated by the numbers in the upper panel (energy density) and the lower panel (energy in the shell) of Figure \ref{Fig02} differ by one.}

\textcolor{black}{Another minor feature of the shell approach is the instability of the numerical process with respect to the input data, see, e.g., \cite{plunian2013shell} for more details. In other words, individual realizations diverge from initially similar distributions, so the average over all realizations is more meaningful. Thus, we add a small noise into the input data (the colored dots in the bottom panel of Figure \ref{Fig02}). The required 500 different sets of initial $|B_n|^2$ and $|U_n|^2$ (for 10 consecutive runs of 50 parallel threads) are obtained by adding 1 percent random noise to each cell energy (they are marked with gray and black lines on the bottom panel, but as the noise is small, they merge into two lines).}

%%%%%%%%%%%%%%%%%%%%%%%%%%%%%%%%%%%%%%%%%%%%%%%
\begin{figure*}
\includegraphics[width = 0.98\textwidth]{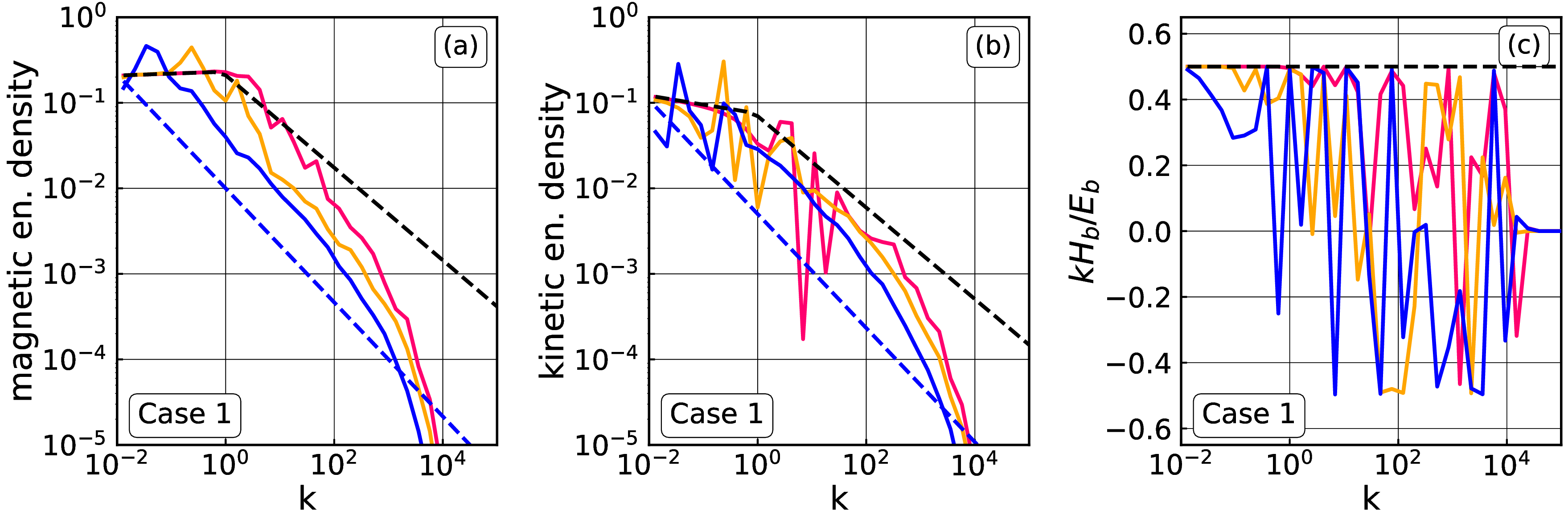}
\includegraphics[width = 0.98\textwidth]{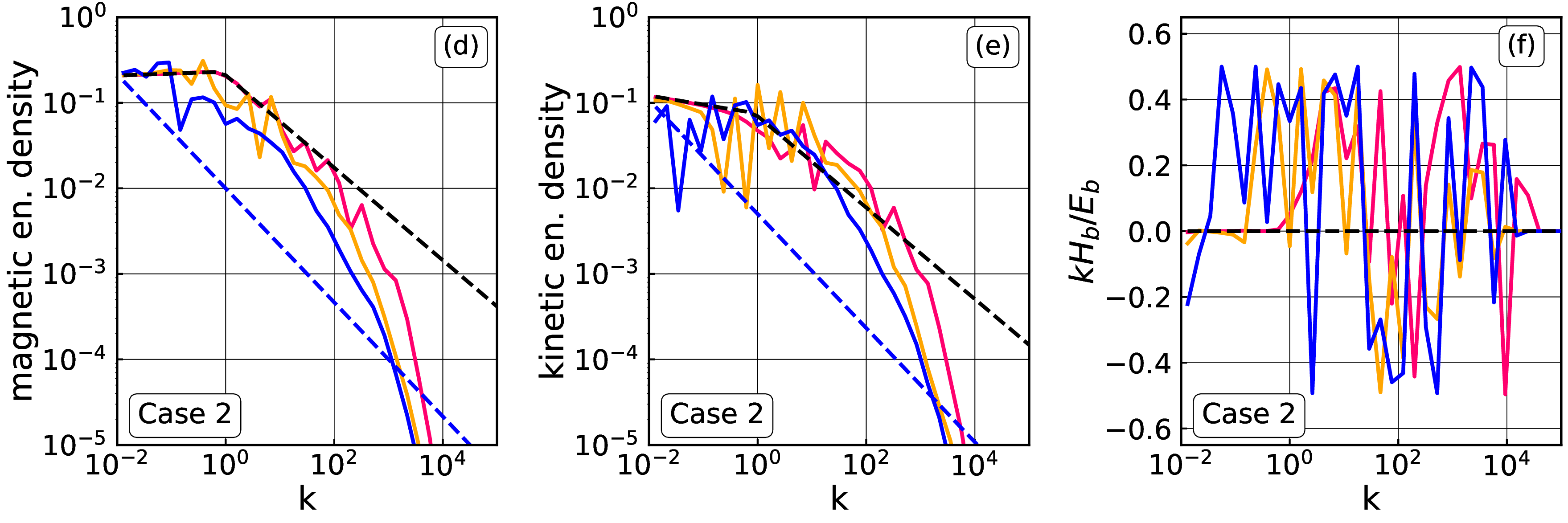}
\caption{The evolution of magnetic energy spectra (left columns, (a,d)), kinetic energy spectra (middle columns, (b,e)), and the ratio $kH_b/E_b$ (right columns, (c,f)) for the moments $t=0.5$ (red lines), $t=5$ (orange lines), and $t=30$ (blue lines). The top row corresponds to the evolution of spectra with maximum initial magnetic helicity $\gamma = 0$ and $\delta = 1$ (Case 1). The bottom row corresponds to the evolution of spectra with zero initial magnetic helicity $\delta = 0$, $\gamma = 1$ (Case 2). In the panels (a,d) and (b,e), the distributions are averaged over $500$ realizations; in the panels (c,f), one particular realization is demonstrated.} 
\label{Fig03}
\end{figure*}
%%%%%%%%%%%%%%%%%%%%%%%%%%%%%%%%%%%%%%%%%%%%%%%

Finally, using these distributions, we construct the real and imaginary components of the collective variables. For this, we introduce two parameters, $\gamma$ and $\delta$, which define the proportionality between the real and imaginary parts of the velocity and magnetic field components in each shell: $b_n = \gamma a_n$ and $c_n = \delta d_n$. So the magnetic helicity and cross helicity take the form:
\begin{equation}\label{Eq05}
\frac{kH_b}{E_b} = \frac{\delta}{1+\delta^2},\quad
\frac{H_c}{\sqrt{E_bE_u} } = \frac{\gamma+\delta}{\sqrt{1+\gamma^2}\sqrt{1+\delta^2}}.
\end{equation}
Varying the input parameters of the shell model, $\gamma$ and $\delta$, leads to different scenarios of energy spectrum evolution for different helicity states. Our choice of these parameters is generally determined by observations. Various experimental studies, such as, e.g., \cite{bavassano1998cross} or \cite{chen2013residual}, describe integral SW characteristics:
\begin{equation}
\sigma_r = \frac{E_u-E_b}{E_u + E_b}, \; \; \; \sigma_c = \frac{2H_c}{E_u + E_b}
\end{equation}
at various radial distances. \textcolor{black}{For example, in the near-solar region, observations show that $\sigma_r$ is mainly negative with the average value of $\langle \sigma_r \rangle = -0.4$. Our non-dimensionalized data give $E_u = 1.2$ and $E_b = 2.9$ for which $\sigma_r = -0.4$ as in observations. The value of $\sigma_c$ can take both positive and negative values in the range $[-1,1]$ with the average $\langle \sigma_c \rangle = 0.6$, e.g., \cite{bruno2013solar,bavassano1998cross}. Our input data gives $\sigma_c = 0.9 (\gamma + \delta) / \sqrt{(1 + \gamma^2) (1 + \delta^2)}$. Thus, it remains within the required range. Moreover, for the demonstrated cases we usually chose one parameter equal to 1 and the other to 0 (see examples below) obtaining $\sigma_c = 0.6$, which also corresponds to the observations. Finally, let us look at the measurement results of the magnetic helicity density by spacecraft (for example, \cite{brandenburg2011scale}).  They show that in individual realizations, the magnetic helicity can be positive or negative with equal probability, so that its average value is close to zero. However, for the absolute value, these observations show a proportionality between the helicity $|kH_b|$ and the energy $E_b$. Thus, in our model cases we choose $\delta=0$ to describe the averaged state or $|\delta| = 1$ to ensure that the proportionality  $|kH_b| = E_b/2$ holds.}

\section{Results of shell modeling}
Numerical simulations based on the shell model were performed for different combinations of both Reynolds numbers within the range $10^4 \leq \Ru,\Rm\leq 10^7$. The simulations showed that a pronounced inertial range (at least two orders of magnitude in $k$-space) does not appear at Reynolds numbers below $10^5$. The higher the Reynolds number, the longer the inertial range is ($k_{\rm dis}/k_{\rm en}\sim\Ru^{3/4}$), but the results do not change qualitatively. Therefore, for the purpose of clarity, we will present all the figures below for ${\Ru}={\Rm}=10^6$. Thus, the only governing parameters are $\delta$ and $\gamma$, which are responsible for the assumed mirror asymmetry at the initial time moment.

Analyzing the spectral density of helicity measured in spacecraft experiments, see, e.g., \cite{brandenburg2011scale, podesta2011magnetic},  one can see that the absolute value of $|kH_b(k)|$ is proportional to the magnetic energy $E_b(k)$ over a wide range of wave numbers. Since this ratio $|kH_b(k)|/E_b(k)$ near the Sun reaches the values of $0.5$, it means from (\ref{Eq05}) that we can choose $\delta=1$ (let us call it Case 1). On the other hand, only the absolute value $|kH_b(k)|$ behaves this way, and the value $kH_b(k)$ behaves quasi-chaotically taking positive and negative values at neighboring points of the spectrum, so the value averaged over each shell turns out to be close to zero. This suggests that we should take $\delta=0$ (let us call it Case 2). Further we will consider both cases in details.

{\bf Case 1 (magnetic helicity of one sign)}. The top panels of Figure \ref{Fig03} show the spectra evolution for $\delta = 1$ ($kH_b/E_b=0.5$) and $\gamma=0$ ($\sigma_c=0.6$). Different colors correspond to different time moments as indicated in the caption. The left and middle panels show the spectral densities of magnetic and kinetic energy, respectively, averaged over 500 realizations. The right panel shows the evolution of $kH_b(k)/E_b(k)$ only for one particular realization. It can be seen that the magnetic energy spectrum exhibits strong variations not only in separated shells but even in the spectral slope. Starting at the power law close to $k^{-3/2}$, the magnetic energy spectrum becomes very steep already at $t=0.5$
(like $k^{-2}$, see the red line in Figure ~\ref{Fig03},a). As it evolves, it becomes smooth and tends towards the spectrum with the slope of $-5/3$ (blue  line). At the same time, a hump forms on the large scales to the left of the inertial range. This hump is not observed in spacecraft data, see Figure \ref{Fig01}, but has been discussed in previous shell simulations of free-decaying MHD turbulence -- if the initial magnetic field possesses a significant helicity, which cannot cascade to small scales, then the magnetic energy associated with this helical field condenses on large scales \cite{Frick2010}. At the early stages of evolution, strong energy fluctuations occur in the averaged kinetic spectrum. The right panel of Figure ~\ref{Fig03} helps to understand where these fluctuations come from showing that the positive helicity imposed on the system at the initial moment of time persists only on large scales, while at $k>5$, already at $t\approx 0.5$, strong fluctuations in helicity occur with a change in sign.
Thus, the evolution of helicity in individual realizations indicates that the initially monotonic distribution of $kH_b(k)/E_b(k)=0.5$ starts to become chaotic. This chaos spreads from small scales to larger ones. Based on this, it can be concluded that the monotonic helicity distribution $kH_b(k)/E_b(k)=0.5$ is not natural for SW evolution, and the development of helicity noise leads to a sawtooth-like structure in the magnetic and kinetic spectra.

{\bf Case 2 (zero mean magnetic helicity)}. The bottom panels of Figure \ref{Fig03} show the spectra evolution for $\delta=0$ ($kH_b/E_b = 0$) and $\gamma = 1$ ($\sigma_c = 0.6$). The results of these simulations (which are all similar to those in the upper panels) demonstrate that, in the case of the average magnetic helicity of zero, a pronounced inertial range does not appear. There are no power-law ranges in the spectral density of either magnetic or kinetic energy; rather, an exponential decrease can be  observed. Even a decrease in dissipation (by the Reynolds numbers growth) does not lead to the formation of a power law range. The spectral distributions also exhibit some sawtooth-like structure, which can be explained by an individual realization shown in the right panel. This panel demonstrates that, despite the conservation of zero magnetic helicity (integrated value), $kH_b/E_b$ begins to become chaotic, gradually alternating in sign between neighboring shells. Over time, this randomization propagates from small to larger scales, while the value $|kH_b(k)/E_b(k)|$ in each shell increases to a possible maximum, approximately equal to $0.5$.

Comparing the results of spacecraft observations with the results of shell modeling, we can suggest that the natural distribution of the spectral helicity can really be non-monotonic but rather saw-toothed, with values of different signs in neighboring wave numbers. This raises questions about whether it is appropriate to use the Bochner-Khinchin theorem to determine the helicity density, or whether the Fourier transform converges and is correct for such a density, or how it relates to the correlation tensor.
However, if we accept this as given, we can still conduct modeling by setting the helicity to be quasi-random at the initial moment. In the third case below, we randomly assign the value $\pm1$ to the $\delta$ in each spectral shell.

%%%%%%%%%%%%%%%%%%%%%%%%%%%%%%%%%%%%%%%%%%%%%%
\begin{figure*}[t]
\centering
\includegraphics[width = 0.32\textwidth]{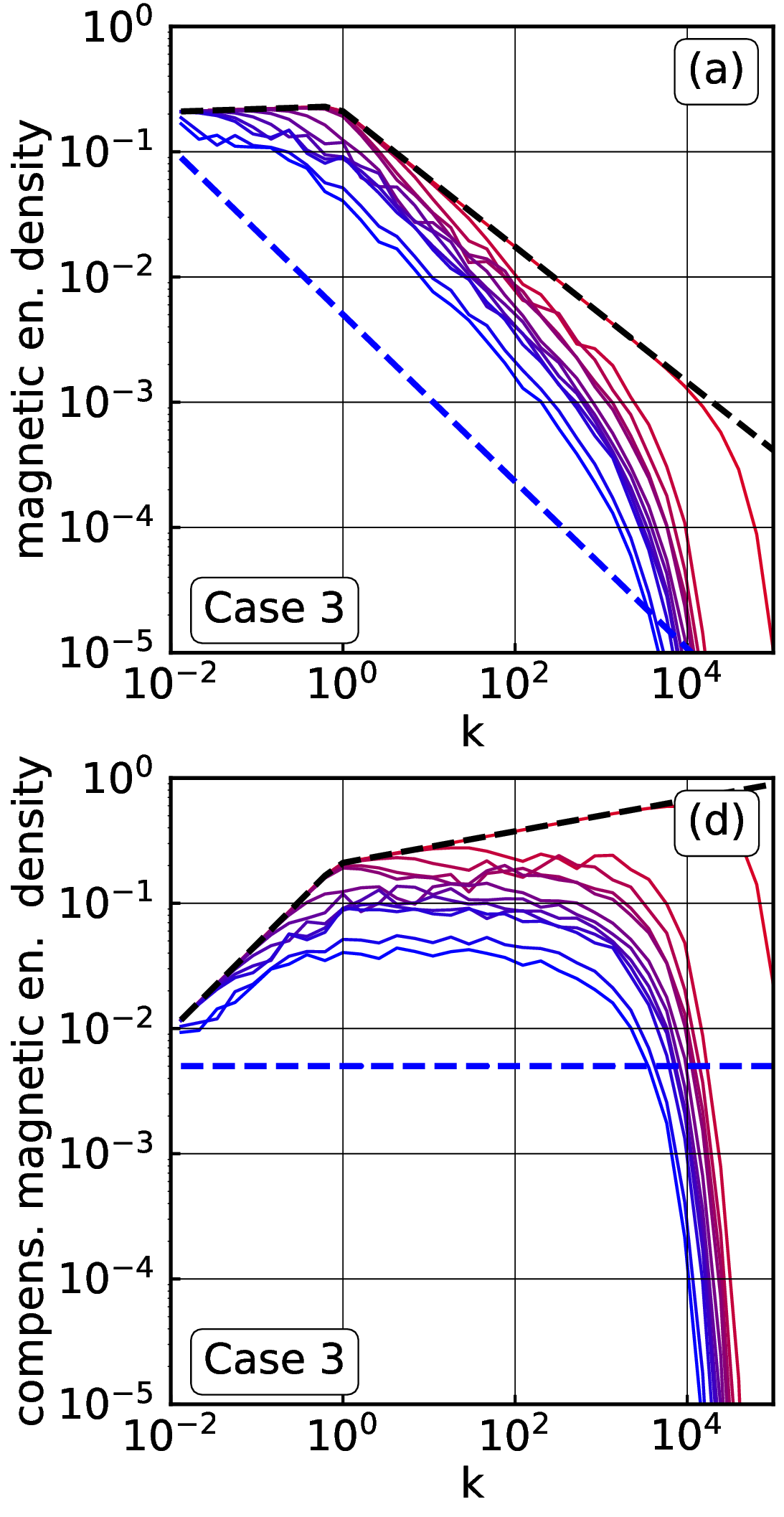}
\includegraphics[width = 0.32\textwidth]{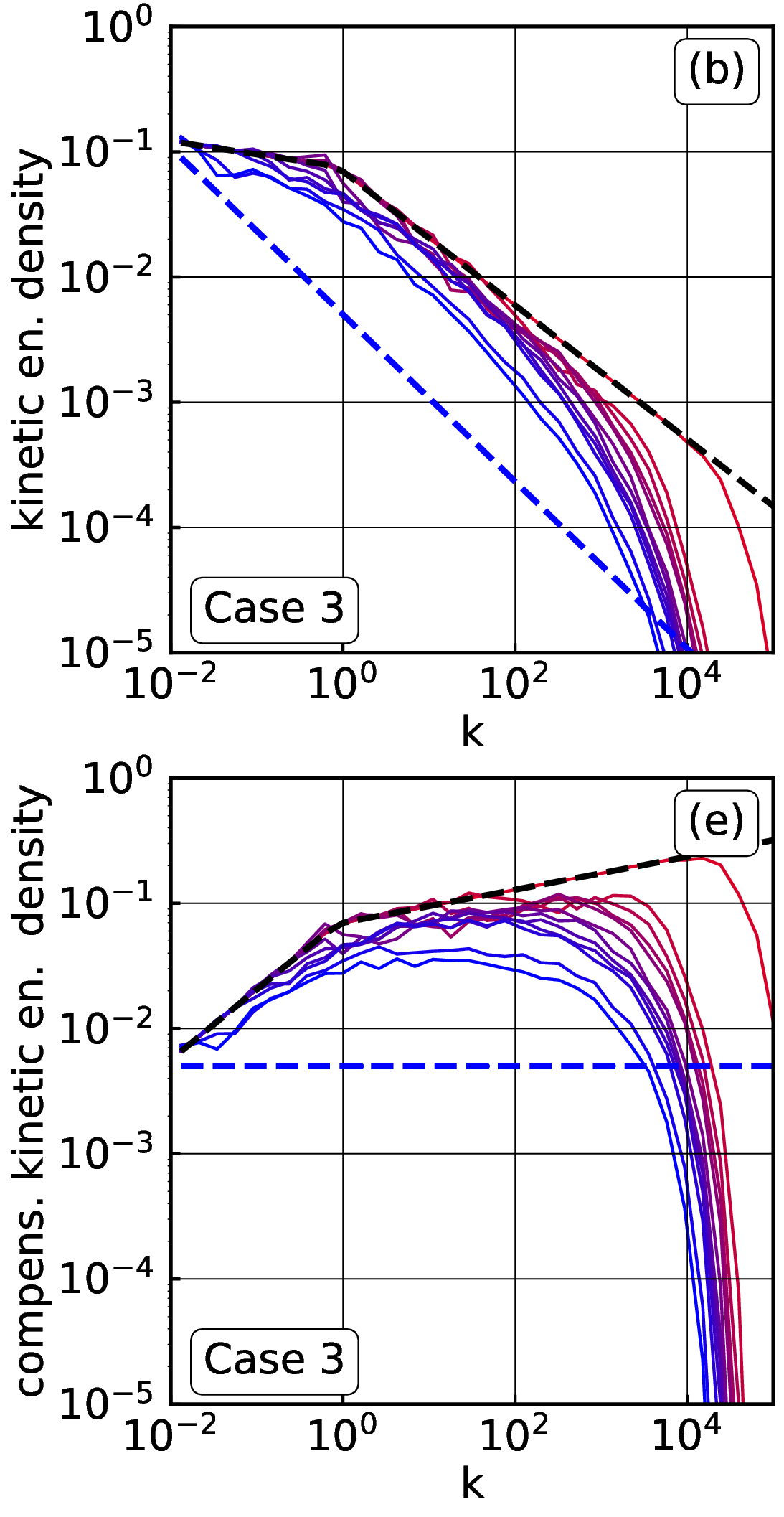}  
\includegraphics[width = 0.324\textwidth]{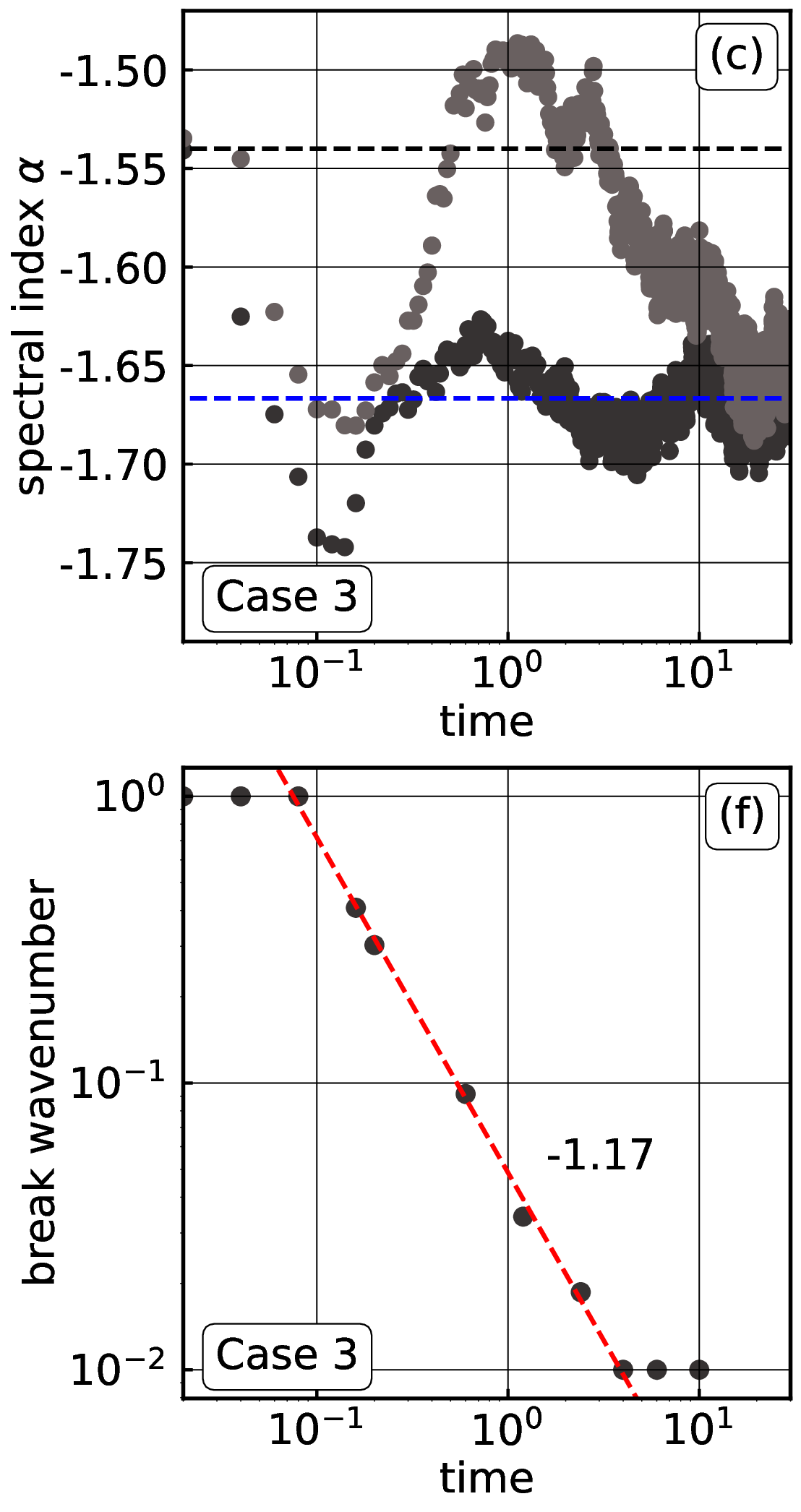}  
\centering
\caption{The evolution of energy density spectra (a,b) and compensated energy spectra (d,e) for Case 3. The left panels (a,d) correspond to the magnetic field and the middle panels (b,e) correspond to the velocity field. The right panels show the change over time of the inertial range slope (spectral index, panel (c)) \textcolor{black}{for magnetic/kinetic energy (black/gray dots)} and the position of the large-scale break in the magnetic energy density spectrum (panel (f)). The time is marked by a gradient from red ($t=0$) to blue ($t=30$), the dashed lines correspond to power-law slopes; $\gamma=0$, $\delta$ is set quasi-chaotically, $\pm1$ in each shell.}
\label{Fig04}
\end{figure*}
%%%%%%%%%%%%%%%%%%%%%%%%%%%%%%%%%%%%%%%%%%%%%%

{\bf Case 3 (magnetic helicity with random sign)}. Figure 4 shows the time evolution of the magnetic (top panel) and kinetic (bottom panel) spectra, from the red (t=0) to the blue (t=30) color. This evolution is typical for the realizations with the random distribution of $\delta$  across the spectrum. Now, at $t=0$  in each given cell the parameter $\delta$ takes values of $\pm1$ with equal probability. Similar results were also obtained for $\delta=\pm0.5$ and for $\delta$ slightly biased towards positive or negative values. In general, this random distribution of the sign of magnetic helicity around the cells allows us to reproduce all the main features of SW spectra evolution observed in reality.

\textcolor{black}{The spectra evolve smoothly, with a change in slope and in the movement of the large-scale break. A similar evolution can be seen in Figure \ref{Fig01}. For the comparison with experimental data, we show not only the magnetic/kinetic energy distributions across the shells (left column), but also the compensated spectra (middle column) similar to the bottom panel of Figure \ref{Fig01}. The latter show corresponding distributions multiplied by $k^{5/3}$ and better demonstrate how rapidly the slope of the spectral density changes. They also show that the energy decays insufficiently fast ($t=30$ corresponds to approximately $R=5$ au), which can be explained by the absence of cooling in the model used.}

In all three cases, averaging over 500 realizations was used, but only in Case 3 the obtained energy spectra are smooth enough to calculate the spectral index $\alpha$ and follow the dynamics of large-scale breaks from them, see the right column of Figure \ref{Fig04}. The spectral index is calculated by approximating a part of the spectrum in the range $k\in[10^0,10^2]$ with a power function. \textcolor{black}{The dynamics of this angle consists of two stages. The first short stage includes $t < 0.1$, where $\alpha$ decreases from $-1.5$ to $-1.75$ for the magnetic energy (black dots) and to $-1.66$ for the kinetic energy (gray dots). The reason for this is that at short times, the most noticeable influence of the turbulent cascade and dissipation is observed on small scales. This causes small-scale energy to drop sharply leading to an increase in the slope of the spectrum. After $t > 0.1$, interactions begin to affect larger scales too, causing the spectral indices to grow up back and then change more smoothly. The slope of the kinetic spectrum returns to $-3/2$, while the spectral index of the magnetic field oscillates near the Kolmogorov slope $-5/3$. This is in line with the spacecraft observations, which show that the spectral index for the magnetic field can reach $-5/3$ near the Sun, \cite{chen2013residual}, and then oscillate around this value \cite{lotz2023radial}. For the velocity field, the slope changes slowly and stabilizes at $-5/3$ only over several au (our estimated time $t=10$-$20$, which corresponds to $1.5$-$3$ au), such behavior is also consistent with the observations, see, e.g., \cite{roberts2010evolution}.} The dynamic of the large-scale break is shown in the bottom panel of Figure \ref{Fig04}. The break point is manually determined by the magnetic field as the point where the inertial interval deviates from the $-1$ slope. The obtained behavior can also be divided into several parts. In the first part, for $t<0.1$, the nonlinear interactions have no effect on the large scales, so the breaking point does not move much. In the last part, after $t>5$, it also does not move because it has reached the left numerical grid boundary. \textcolor{black}{However, in the interval between these moments, there is a clear power-law motion of the point towards larger scales, with $k_b\sim t^{-1.17}$, which is in excellent agreement with the observational data, see, e.g., \cite{chen2020evolution}, that estimate the motion of this point according to the law $k_b\sim t^{-1.25\pm0.11}$. Here we make a comparison with a study in which this dependence is calculated for the plasma travel time from the Sun. There are also many papers in which similar power-law dependencies are calculated for distances $R$, e.g., \cite{lotz2023radial}. As noted above, the conversion of travel time to distance can be roughly estimated by multiplying by $U_f$. However, a detailed comparison requires taking into account the increase in plasma velocity with distance from the Sun.}

\begin{figure}[t]
\centering
\includegraphics[width = 0.47\textwidth]{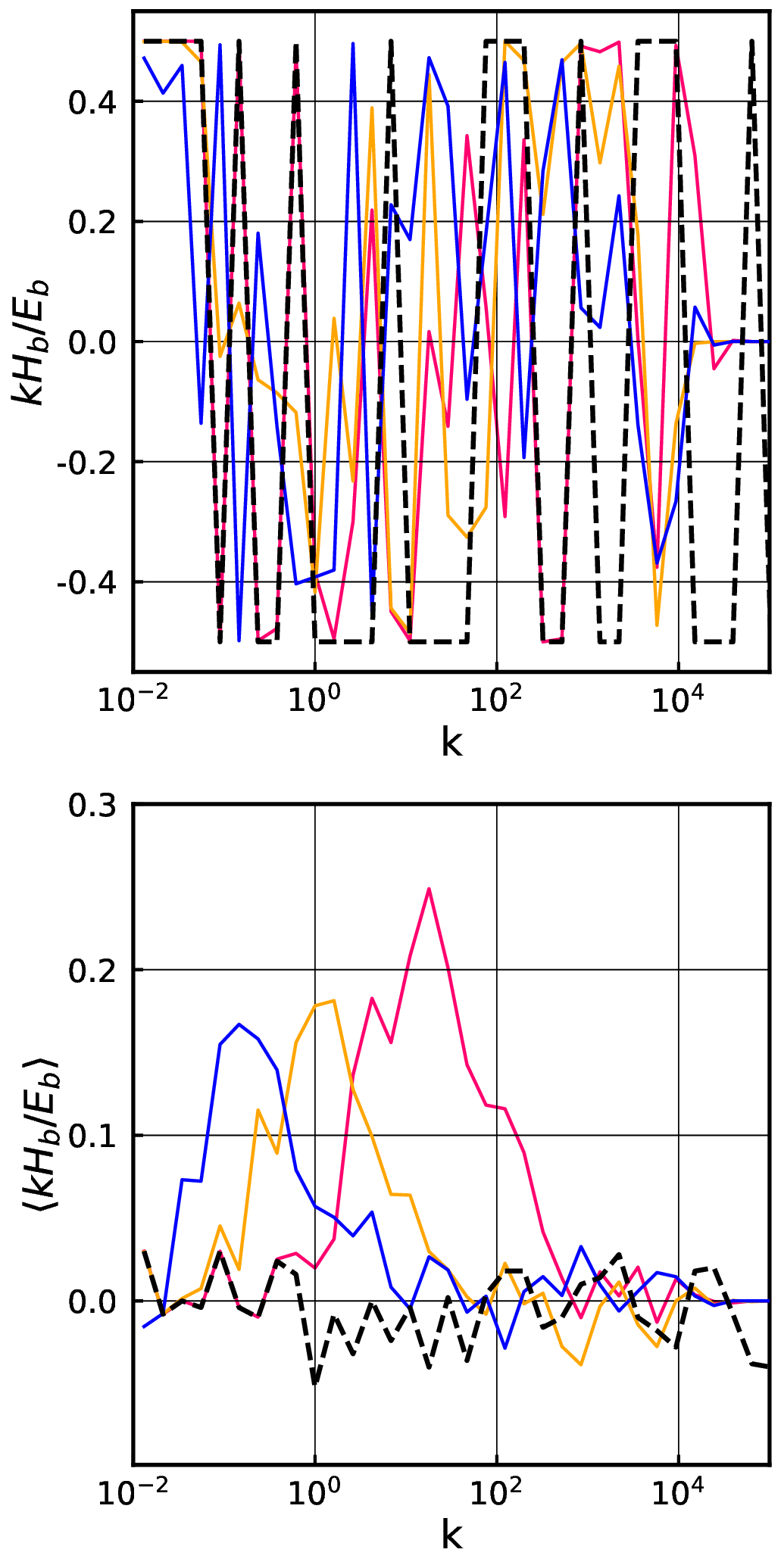}  
\centering
\caption{Spectral distribution of the ratio of magnetic helicity to magnetic energy $kH_b/E_b$. The top panel shows an individual realization (dashed line) and the average over all realizations (solid line). The bottom panel shows the dynamics of the average distributions of $kH_b/E_b$ for different time instances: $t=0.5$ (red line), $t=5$ (yellow line) and $t=30$ (blue line).}
\label{Fig05}
\end{figure}

Finally, we note that the spectral distribution of magnetic helicity, defined as a quasi-chaotic at the initial moment, remains so over time. We understand the mathematical questions of such a Fourier image, however, as mentioned earlier, this fits perfectly with the observational results \cite{brandenburg2011scale}. Figure \ref{Fig05} shows the dynamics of $kH_b/E_b$ for different time points, from which the quasi-chaotic behavior of sawtooth structure can be clearly seen. However, this behavior of individual realizations does not mean that the average value is not monotonic. In the lower panel, the average value of $kH_b/E_b$ over 500 realizations shows that magnetic helicity starts to be dominated by one sign in the large-scale part of the spectrum over time. A hump forms in the average value that shifts to larger scales over time. At the same time, the integral helicity remains close to zero. In the discussion, we will comment on the accumulation of helicity at large scales in more detail. However, it should be highlighted here that this accumulation is stable with respect to the initial distributions. Of course, the sign of the helicity depends on the definition of its integral: $H_b=\sum c_nd_n/k$ or $H_b=-\sum c_nd_n/k$. Here we choose the first case, so we always observe the positive helicity accumulation. The second one gives us negative helicity that is better consistent with the observational data, which show a negative dome-shaped structure in the average $kH_b/E_b$ (see \cite{brandenburg2011scale}), as well as the cascade modeling data, which show a negative spiral flow in previous studies, e.g. \cite{stepanov2014joint}. 

\textcolor{black}{Considering the third conserved in ideal MHD quadratic quantity, the cross-helicity, we have to remember that we study a free decay of MHD turbulence without any injection of energy and helicities. Thus, since the average cross-helicity and magnetic helicity were initially close to zero, they remained small throughout the entire time interval considered. This choice was motivated by spacecraft observations showing the presence of both positive and negative helicity values in the solar wind. The cross-helicity can affect the energy cascade, as it was shown by shell models (e.g., \cite{mizeva2009cross}) and should be involved in future models together with the description of the expansion and cooling of the solar wind and its anisotropy. The shell modeling presented here demonstrates that, in general, all of this is feasible, but each of these issues will need to be considered carefully and separately.}

\section{Discussion and conclusion}
The idea of using a shell model to describe the evolution of turbulent spectra emerged nearly half a century ago. This significantly simplified the MHD description but left us with questions: To what extent were physical effects lost in this simplification? In other words, what physical effects are still present due to solely quadratic nonlinearities and corresponding conservation laws? To address this question, we are undertaking another attempt to model the evolution of the turbulent spectra of the solar wind. Our goal is twofold: first, to track the main features of the dynamics of the large-scale MHD part of the energy spectrum, and second, to obtain estimates for the parameters such as helicity that are easy to vary in a simplified model, but difficult to extract from spacecraft observations.

This attempt is not the first, and it is likely not the last, to describe SW dynamics using the shell approach, see, e.g., \cite{carbone2002extent,consolini2015emergence,verdini2012origin}. A significant feature of our approach is the use of a mirror-asymmetric model with helicity, which is more flexible than in most other shell models of the GOY type \cite{mizeva2009cross,mizeva2009role}. Specifically, the sign of helicity can be arbitrary, and there is no strict proportionality between energy and helicity for each specific shell. Furthermore,  the spacecraft data available after the launch of the PSP mission allow the initial spectra to be set at much closer distances to the Sun, see, e.g., \cite{raouafi2023parker}. As a result, at small distances of the order of 1 au (where the thermodynamic effects related to SW expansion and energy/helicity injection at kinetic scales can be ignored), it becomes possible to simulate the energy/helicity dynamics over a longer period of time.

Here, we again should note that all the results presented in the paper were obtained for Reynolds numbers ${\rm Re}=10^6$ and ${\rm Rm}=10^6$. The shell modeling demonstrated that, for higher Re and Rm, it is not the fundamental structure or dynamics that changes but rather the length of the inertial range. At lower Reynolds numbers, an adequate inertial range was not formed (in terms of two orders of the magnitude of wavenumbers compared to the observational data). It is important that this does not conflict with the existing estimates as this range includes the proposed numbers as a particular case \cite{wrench2024reynolds}. The choice of Reynolds number determined the formulation of the problem as a free energy degeneracy meaning that there was no energy pumping on either small or large scales, and the cooling of the solar wind due to plasma expansion was not considered. Energy monotonically decreased during evolution, transferred at a small scale and then dissipated. Simultaneously, the evolution time was chosen to be quite short, $t<30$. Therefore, in the context of applying the results to the real solar wind, they can only be relied upon in the immediate vicinity of the Sun, up to a few astronomical units.

%%%%%%%%%%%%%%%%%%%%%%%%%%%%%%%%%%%%%%%%%%%%%%
%%%%%%%%%%%%%%%%%%%%%%%%%%%%%%%%%%%%%%%%%%%%%%

The main factor in the spectrum evolution turned out to be the setting of the mirror asymmetry of the input data for shell modeling. It was found that, when the spectral helicity is zero or maximum within the model framework, the dynamics of the energy begins with spectral randomization. In other words, the monotonous spectral distribution of helicity becomes sawtooth-shaped gradually covering smaller to larger scales. Since the conservation laws are inherent in the model, the transition to a sawtooth shape preserves integral value: the helicities in neighboring shells simply become opposite in sign. This behavior can be attributed to a specific feature of the model. Nevertheless, the significance of the result is attributed to the fact that a similar pattern has been observed in real solar wind plasma according to the spacecraft data. Thus, we can conclude that the concept of spectral helicity density may not be applicable to individual implementations, or it should be understood as an average over many implementations. This chaotic evolution results in a noisy structure for the energy-averaged spectra, and their evolution significantly differs from realistic behavior. Therefore, we can use non-monotonic and chaotic spectral realizations as input data, which we believe to be correct and natural.

The initial spectral distribution of helicity with a random sign in each spectral shell leads to the realistic evolution of energy spectra averaged over all the realizations. During this evolution, we observe a smooth transfer of magnetic energy to kinetic energy, an inertial range forms, the spectral index transforms to $-5/3$, and the large-scale break shifts to larger scales. Interestingly, spectral indices tend towards the Kolmogorov slope of $-5/3$ in different ways for both the magnetic field and the velocity field, compare, e.g., with \cite{roberts2010evolution}. The magnetic spectrum achieves the Kolmogorov slope almost immediately fluctuating around it. After a slight adjustment, the kinetic spectrum returns to the $-3/2$ slope and then relaxes to $-5/3$ over a long period of time. In terms of distance, this corresponds to approximately a few au. The breaking point has moved along the spectrum over the same long period clearly following the power law $k_b\sim t^{-7/6}$, compare, e.g., with \cite{lotz2023radial}. These behavior markers, despite the simplicity of the shell model, match perfectly with the results of the spacecraft observations.

It is worth to note that, over time, the quasi-chaotic distribution of helicity in the spectrum for each particular realization persists. This confirms the naturalness of this distribution in individual cases. However, what is more interesting, is that the average of these distributions is not chaotic or even zero. Instead, it is quite monotonous with the clear maximum of the ratio $kH_b/E_b$. Although this maximum is not present at the beginning of the process, it forms and moves from the dissipative region towards a large-scale break. Unexpectedly, we have found a mechanism that could lead to the violation of symmetry between right and left. This phenomenon is widely discussed in particle physics and the study of the Early Universe, in particular in the context of MHD processes, e.g. \cite{Ber25}. In this paper, we investigated the evolution of the magnetic field in the solar wind. It would be inappropriate to directly compare our results with those from the study of the Early Universe, but we believe the identified mechanism could be interesting in that context.

The last thing we would like to mention is that the accumulation of the average magnetic helicity at a large scale has also been observed in experimental data, see, e.g., \cite{brandenburg2011scale}. The sign of the $kH_b/E_b$ hump appears to be different, but, as we have already discussed, this is not fundamental, because it can be corrected by choosing a different sign for the conservation law in the shell model. The reason for this accumulation is not completely understood, and it is possible that it may be related to the negative flow of positive helicity or a slight asymmetry in the background hydrodynamic mirror-asymmetry. The obtained simulation results make it even more interesting to pose various questions during future spacecraft missions. The questions may be as follows: Whether the accumulation in the northern and southern hemispheres of the solar system differs? Whether this $kH_b/E_b$ hump is moving towards a larger scale over time? Whether such magnetic helicity accumulation is observed in other turbulent plasma flows or only in the solar wind?

\section{Acknowledgments}
The authors would like to express their sincere gratitude to A. Vinogradov and H. Malova for their invaluable assistance in analyzing the extensive literature on the subject of the solar wind and for their constant readiness to help understand the issues related to its dynamics. We would also like to express our sincere gratitude to everyone who contributed to the PSP project. In particular, we would like to thank the FIELDS and SWEAP teams for their valuable data that were used in this study (all data are publicly  available via the NASA Space Physics Data Facility, https://spdf.gsfc.nasa.gov/).  P.F. acknowledges the support of RAS under the project 124012300246-9.

\section*{References}
\bibliography{IliaBib}

@article{alexandrova2013solar,
  title={Solar wind turbulence and the role of ion instabilities},
  author={Alexandrova, O. and Chen, C. and Sorriso-Valvo, L. and Horbury, T. S. and Bale, S. D.},
  journal={Space Science Reviews},
  volume={178},
  number={2},
  pages={101--139},
  year={2013},
  publisher={Springer}
}

@article{bandyopadhyay2020direct,
  title={Direct measurement of the solar-wind Taylor microscale using MMS turbulence campaign data},
  author={Bandyopadhyay, R. and Matthaeus, W. H. and Chasapis, A. and Russell, C. T. and Strangeway, R. J. and Torbert, R. B. and Giles, B. L. and Gershman, D. J. and Pollock, C. J. and Burch, J. L.},
  journal={The Astrophysical Journal},
  volume={899},
  number={1},
  pages={63},
  year={2020},
  publisher={IOP Publishing},
  doi={10.3847/1538-4357/ab9ebe}
}

@ARTICLE{Ber25,
       author = {Bershadskii, A.},
        title = "{Galactic foreground and CMB emissions randomization due to chaotic/turbulent dynamics of magnetized plasma dominated by magnetic helicity}",
      journal = {arXiv e-prints},
         year = 2025,
        month = feb,
          eid = {arXiv:2502.18379},
        pages = {arXiv:2502.18379},
          doi = {10.48550/arXiv.2502.18379},
archivePrefix = {arXiv},
       eprint = {2502.18379},
 primaryClass = {astro-ph.GA},
       adsurl = {https://ui.adsabs.harvard.edu/abs/2025arXiv250218379B}
}

@misc{Bale_MacDowal_Koval_Pulupa_Quinn_Schroeder_2020, title={{PSP FIELDS Fluxgate Magnetometer (MAG) Magnetic Field Vectors, Radial-Tangential-Normal, RTN, Coordinates, 4 samples/cycle, Level 2 (L2), 3.413 ms Data}}, url={https://spase-metadata.org/NASA/NumericalData/ParkerSolarProbe/FIELDS/MAG/Level2/RTN/4SamplesPerCycle/PT0.003413S}, DOI={10.48322/WQCS-A534}, abstractNote={Parker Solar Probe FIELDS Instrument Suite Fluxgate Magnetometer, MAG, Data: The time resolution of the MAG time series data varies with instrument mode ranging from 2.289 samples/s to 292.9 samples/s. These two data sampling rates corresponding to 2 samples or 256 samples per 0.874 s where 0.874 s is equal to 2^25 divided 38.4 MHz. See reference [2] for a complete explanation of the MAG instrument sampling methodology. The Magnetometer has four ranges: ±1024 nT, ±4096 nT, ±16,384 nT, and ±65,536 nT. The Magnetometer Range is selected by an algorithm based on the strength of the ambient magnetic field. The magnetic field measurement precision is ±15 bits, based on the 16-bit Analog to Digital Converter, ADC.
References:
* 1. Fox, N.J., Velli, M.C., Bale, S.D. et al. Space Sci Rev (2016) 204: 7. https://doi.org/10.1007/s11214-015-0211-6
* 2. Bale, S.D., Goetz, K., Harvey, P.R. et al. Space Sci Rev (2016) 204: 49. https://doi.org/10.1007/s11214-016-0244-5}, publisher={Space Physics Data Facility}, author={Bale, S. D. and MacDowal, R. J. and Koval, A. and Pulupa, M. and Quinn, T. and Schroeder, P.}, year={2020}, language={en} }

@article{bavassano1998cross,
  title={Cross-helicity and residual energy in solar wind turbulence: Radial evolution and latitudinal dependence in the region from 1 to 5 au},
  author={Bavassano, B. and Pietropaolo, E. and Bruno, R.},
  journal={Journal of Geophysical Research: Space Physics},
  volume={103},
  pages={6521--6529},
  year={1998},
  publisher={Wiley Online Library},
  doi={10.1029/97JA03029}
}

@article{brandenburg2011scale,
  title={Scale dependence of magnetic helicity in the solar wind},
  author={Brandenburg, A. and Subramanian, K. and Balogh, A. and Goldstein, M. L.},
  journal={The Astrophysical Journal},
  volume={734},
  pages={9},
  year={2011},
  publisher={IOP Publishing},
  doi={10.1088/0004-637X/734/1/9}
}

@article{brandenburg2023turbulence,
  title={Turbulence with magnetic helicity that is absent on average},
  author={Brandenburg, A. and Larsson, G.},
  journal={Atmosphere},
  volume={14},
  number={6},
  pages={932},
  year={2023},
  publisher={MDPI},
  doi={10.3390/atmos14060932}
}

@article{bruno2013solar,
  title={The solar wind as a turbulence laboratory},
  author={Bruno, R. and  Carbone, V.},
  journal={Living Reviews in Solar Physics},
  volume={10},
  pages={2},
  year={2013},
  publisher={Springer},
  doi={10.12942/lrsp-2013-2}
}

@book{bruno2016turbulence,
  title={Turbulence in the solar wind},
  author={Bruno, R. and Carbone, V.},
  volume={928},
  year={2016},
  publisher={Springer},
  doi={10.1007/978-3-319-43440-7}
}

@article{carbone2002extent,
  title={To what extent can dynamical models describe statistical features of turbulent flows?},
  author={Carbone, V. and Cavazzana, R. and Antoni, V. and Sorriso-Valvo, L. and Spada, E. and Regnoli, G. and Giuliani, P. and Vianello, N. and Lepreti, F. and Bruno, R. and others},
  journal={Europhysics letters},
  volume={58},
  number={3},
  pages={349},
  year={2002},
  publisher={IOP Publishing},
  doi={10.1209/epl/i2002-00645-y}
}

@article{chen2013residual,
  title={Residual energy spectrum of solar wind turbulence},
  author={Chen, C. H. K. and Bale, S. D. and Salem, C. S. and Maruca, B. A.},
  journal={The Astrophysical Journal},
  volume={770},
  pages={125},
  year={2013},
  publisher={IOP Publishing},
  doi={10.1088/0004-637X/770/2/125 }
}

@article{chen2020evolution,
  title={The evolution and role of solar wind turbulence in the inner heliosphere},
  author={Chen, C. H. K. and Bale, S. D. and others},
  journal={The Astrophysical Journal Supplement Series},
  volume={246},
  pages={53},
  year={2020},
  publisher={IOP Publishing},
  doi={10.3847/1538-4365/ab60a3}
}

@article{consolini2015emergence,
  title={On the emergence of a 1/k spectrum in the sub-inertial domain of turbulent media},
  author={Consolini, G. and De Marco, R. and Carbone, V.},
  journal={The Astrophysical Journal},
  volume={809},
  number={1},
  pages={21},
  year={2015},
  publisher={IOP Publishing},
  doi={10.1088/0004-637X/809/1/21}
}

@article{cuesta2022intermittency,
  title={Intermittency in the expanding solar wind: Observations from Parker Solar Probe (0.16 au), Helios 1 (0.3--1 au), and voyager 1 (1--10 au)},
  author={Cuesta, M. E. and Parashar, T. N. and Chhiber, R. and Matthaeus, W. H.},
  journal={The Astrophysical Journal Supplement Series},
  volume={259},
  pages={23},
  year={2022},
  publisher={IOP Publishing},
  doi={10.3847/1538-4365/ac45fa}
}

@article{hand2026empirical,
  title={Empirical results on the magnetic Prandtl number in the slow solar wind based on in situ measurements},
  author={Hand, T. J. E. and Roberts, O. W. and Li, X. and Knight, T. and Narita, Y. and V{\"o}r{\"o}s, Z. and Pit{\v{n}}a, A.},
  journal={Journal of Geophysical Research: Space Physics},
  volume={131},
  number={3},
  pages={e2026JA035053},
  year={2026},
  publisher={Wiley Online Library},
  doi={10.1029/2026JA035053}
}

@article{dakeyo2022statistical,
  title={Statistical analysis of the radial evolution of the solar winds between 0.1 and 1 au and their semiempirical isopoly fluid modeling},
  author={Dakeyo, J. B. and Maksimovic, M. and D{\'e}moulin, P. and Halekas, J. and Stevens, M. L.},
  journal={The Astrophysical Journal},
  volume={940},
  number={2},
  pages={130},
  year={2022},
  publisher={The American Astronomical Society},
  doi={10.3847/1538-4357/ac9b14}
}

@book{davidson2017introduction,
  title={Introduction to magnetohydrodynamics},
  author={Davidson, P. A.},
  volume={55},
  year={2017},
  publisher={Cambridge university press}
}

@article{ditlevsen1997cascades,
  title={Cascades of energy and helicity in the {GOY} shell model of turbulence},
  author={Ditlevsen, P. D.},
  journal={Physics of Fluids},
  volume={9},
  number={5},
  pages={1482--1484},
  year={1997},
  publisher={American Institute of Physics},
  doi={10.1063/1.869270}
}

@book{harvey2007soviet,
  title={Soviet and Russian lunar exploration},
  author={Harvey, B.},
  year={2007},
  publisher={Springer},
  doi={10.1007/978-0-387-73976-2}
}

@article{halekas2022radial,
  title={The radial evolution of the solar wind as organized by electron distribution parameters},
  author={Halekas, J. S. and Whittlesey, P. and Larson, D. E. and Maksimovic, M. and Livi, R. and Berthomier, M. and Kasper, J. C. and Case, A. W. and Stevens, M. L. and Bale, S. D. and others},
  journal={The Astrophysical Journal},
  volume={936},
  number={1},
  pages={53},
  year={2022},
  publisher={The American Astronomical Society},
  doi={10.3847/1538-4357/ac85b8}
}

@misc{Kasper_Stevens_Case_Korreck_2020, title={{PSP Solar Wind Electrons Alphas and Protons (SWEAP) SPC Ion Number Density, Velocity, and Thermal Speed Moments and Fits, Level 3 }}, url={https://spase-metadata.org/NASA/NumericalData/ParkerSolarProbe/SWEAP/SPC/Level3/SolarWindMomentsFits/PT0.2185S}, DOI={10.48322/49WE-TR31}, abstractNote={This data product contains derived measurements of ion properties in the solar wind, including those for density, temperature, and velocity. These are determined both by a direct computation of the velocity moments of the reduced distribution function and by attempting to fit the primary peak in the ion I(V) curve with a Maxwellian model. These measurements correspond one to one with spectra in the psp_swp_spc_l2i file for the same date. It may be convenient for some applications to cross-reference the two. For example, the corresponding l3i file contains ephemeris and data quality flag information that may be useful for an investigator who is concerned only with l2i type measurements.}, publisher={NASA Space Physics Data Facility}, author={Kasper, J. C. and Stevens, M. L. and Case, A. W. and Korreck, K. E.}, year={2020} }

@article{lotz2023radial,
  title={The radial variation of the solar wind turbulence spectra near the kinetic break scale from parker solar probe measurements},
  author={Lotz, S. and Nel, A. E. and Wicks, R. T. and Roberts, O. W. and Engelbrecht, N. E. and Strauss, R. D. and Botha, G. and Kontar, E. P. and Pit{\v{n}}a, A. and Bale, S. D.},
  journal={The Astrophysical Journal},
  volume={942},
  number={2},
  pages={93},
  year={2023},
  publisher={IOP Publishing},
  doi={10.3847/1538-4357/aca903}
}

@ARTICLE{Lorenz1972,
  author = {Lorenz, E. N.},
  title = {Low order models representing realizations of turbulence},
  journal = {Journal of Fluid Mechanics},
  year = {1972},
  volume = {55},
  pages = {545-563},
  doi = {10.1017/S0022112072002009},
  url_ = {http://adsabs.harvard.edu/abs/1972JFM....55..545L}
}

@article{markovskii2015statistical,
  title={Statistical analysis of the magnetic helicity signature of the solar wind turbulence at 1 au},
  author={Markovskii, S. A. and Vasquez, B. J. and Smith, C. W.},
  journal={The Astrophysical Journal},
  volume={806},
  pages={78},
  year={2015},
  publisher={IOP Publishing}
}

@article{mcintyre2023properties,
  title={Properties underlying the variation of the magnetic field spectral index in the inner solar wind},
  author={McIntyre, J. R. and Chen, C. H. K. and Larosa, A.},
  journal={The Astrophysical Journal},
  volume={957},
  number={2},
  pages={111},
  year={2023},
  publisher={The American Astronomical Society},
  doi={10.3847/1538-4357/acf3dd}   
}

@article{mizeva2009role,
author = {Stepanov, R. and Frick, P. and Mizeva, I.},
year = {2013},
month = {01},
pages = {15-21},
title = {Cross helicity and magnetic helicity cascades in {MHD} turbulence},
volume = {49},
journal = {Magnetohydrodynamics},
doi = {10.22364/mhd.49.1-2.3}
}

@inproceedings{mizeva2009cross,
  title={The cross-helicity effect on cascade processes in MHD turbulence},
  author={Mizeva, I. A. and Stepanov, R. A. and Frick, P. G.},
  booktitle={Doklady physics},
  volume={54},
  number={2},
  pages={93--97},
  year={2009},
  organization={SP MAIK Nauka/Interperiodica Dordrecht},
  doi={10.1134/S1028335809020128}
}

@article{Obukhov1971,
  author = {Obukhov, A. M.},
  title = {Some general characteristic equations of the dynamics of the atmosphere},
  journal = {Atmos. Oceanic Europhys.},
  year = {1971},
  volume = {7},
  pages = {41},
}

@misc{Papitashvili_2023, title={{BepiColombo Ephemeris, Heliocentric Trajectories, Heliographic, Heliographic Inertial, and Solar Ecliptic Coordinates, HelioWeb, Daily Data}}, url={https://spase-metadata.org/NASA/NumericalData/BepiColombo/HelioWeb/Ephemeris/P1D}, DOI={10.48322/CYEC-7N35}, abstractNote={Heliocentric trajectories for Comet Borrelly in Heliographic, HG, Heliographic Inertial, HGI, and Solar Ecliptic, SE, Coordinates The original trajectory data are taken from http://ssd.jpl.nasa.gov/horizons.cgi where users can find many more objects. In the case of orbit data for planets, the orbit data can be used as a proxy for spacecraft ephemeris that are in orbit about the planets. On a heliospheric scale, differences between the planet orbital tarjectory and that of the spacecraft are very small. For instance, the heliocentric longitudes differ by only 0.25° for a spacecraft stationed near the L1 Lagrange point at approximately 100 Earth radii upstream of the Earth. The production of the HG, HGI, and SE trajectory data requires a values for the “Equinox Epoch”, which is defined as the epoch time when the direction from the Earth to the sun at the time of the vernal equinox when the sun seems to cross equatorial plane of the Earth from below. This direction is called the First Point of Aries, FPA and it is not a fixed direction but drifts by about 1.4° per century or 50.26" per year. In addition, there are tiny irregularities in FPA drift that are on the order of 1" per year or less. The Equinox Epoch can be determined by using a variety of methods for calculating the instantaneous FPA longitudinal direction and whether the tiny irregularities have been smoothed or averaged out. Four methods for determining the Equinox Epoch are in common usage: +---------------------------------------------------------------------+   Method Name   FPA Longitude Definition    ---------------------------------------------------------------------    B1950.0   the actual FPA at 22:09 UT on December 31, 1949     J2000.0   the smoothed FPA at 12:00 UT on January 1, 2000     True of Date   the actual FPA at 00:00 UT on the date of interest     Mean of Date   the smoothed FPA at 00:00 UT on the date of interest   +---------------------------------------------------------------------+ The heliocentric trajectory data included in this data product have been calculated by using the Equinox Epoch: defined via the “Mean of Date” method. More precise coordinates, and some planet-centered coordinates, are found in the “traj” subdirectories of spacecraft specific directories at https://spdf.gsfc.nasa.gov/pub/data/ and http://ssd.jpl.nasa.gov/horizons.cgi.}, publisher={Space Physics Data Facility}, author={Papitashvili, N. E.}, year={2023}, language={en} }

@article{perez2004empirical,
  title={Empirical values of the transport coefficients of the solar wind: Conditions in the Venus ionosheath},
  author={P{\'e}rez-de-Tejada, H},
  journal={The Astrophysical Journal},
  volume={618},
  number={2},
  pages={L145},
  year={2004},
  publisher={IOP Publishing},
  doi={10.1086/425864},
}

@article{phillips2022taylor,
  title={Taylor microscale and effective Reynolds number near the sun from {PSP}},
  author={Phillips, C. and Bandyopadhyay, R. and McComas, D. J.},
  journal={The Astrophysical Journal},
  volume={933},
  pages={33},
  year={2022},
  publisher={IOP Publishing},
  doi={10.3847/1538-4357/ac713f}
}

@article{plunian2013shell,
  title={Shell models of magnetohydrodynamic turbulence},
  author={Plunian, F. and Stepanov, R. and Frick, P.},
  journal={Physics Reports},
  volume={523},
  pages={1--60},
  year={2013},
  publisher={Elsevier},
  doi={10.1016/j.physrep.2012.09.001}
}

@article{podesta2011magnetic,
  title={Magnetic helicity spectrum of solar wind fluctuations as a function of the angle with respect to the local mean magnetic field},
  author={Podesta, J. J. and Gary, S. P.},
  journal={The Astrophysical Journal},
  volume={734},
  pages={15},
  year={2011},
  publisher={IOP Publishing},
  doi={10.1088/0004-637X/734/1/15}
}

@article{raouafi2023parker,
  title={Parker solar probe: Four years of discoveries at solar cycle minimum},
  author={Raouafi, N. E. and Matteini, L. and Squire, J. and Badman, S. T. and Velli, M. and Klein, K. G. and Chen, C. H. K. and Matthaeus, W. H. and Szabo, A. and Linton, M. and others},
  journal={Space Science Reviews},
  volume={219},
  number={1},
  pages={8},
  year={2023},
  publisher={Springer},
  doi={10.1007/s11214-023-00952-4}
}

@article{roberts2010evolution,
  title={Evolution of the spectrum of solar wind velocity fluctuations from 0.3 to 5 AU},
  author={Roberts, D. A.},
  journal={Journal of Geophysical Research: Space Physics},
  volume={115},
  number={A12},
  year={2010},
  publisher={Wiley Online Library}
}

@article{stepanov2014joint,
  title={Joint inverse cascade of magnetic energy and magnetic helicity in MHD turbulence},
  author={Stepanov, R. and Frick, P. and Mizeva, I.},
  journal={The Astrophysical Journal Letters},
  volume={798},
  number={2},
  pages={L35},
  year={2014},
  publisher={IOP Publishing},
  doi={10.1088/2041-8205/798/2/L35}
}

@article{verdini2012origin,
  title={On the origin of the 1/f spectrum in the solar wind magnetic field},
  author={Verdini, A. and Grappin, R. and Pinto, R. and Velli, M.},
  journal={The Astrophysical Journal Letters},
  volume={750},
  number={2},
  pages={L33},
  year={2012},
  publisher={IOP Publishing},
  doi={10.1088/2041-8205/750/2/L33}
}

@article{weygand2009anisotropy,
  title={Anisotropy of the Taylor scale and the correlation scale in plasma sheet and solar wind magnetic field fluctuations},
  author={Weygand, J. M. and Matthaeus, W. H. and Dasso, S. and Kivelson, M. G. and Kistler, L. M. and Mouikis, C.},
  journal={Journal of Geophysical Research: Space Physics},
  volume={114},
  year={2009},
  publisher={Wiley Online Library},
  doi={10.1029/2008JA013766}
}

@ARTICLE{wrench2024reynolds,
    author = {Wrench, D. and Parashar, T. N. and Oughton, S. and de Lange, K. and Frean, M.},
    title = "{What is the Reynolds Number of the Solar Wind?}",
    journal = {The Astrophysical Journal},
    year = 2024,
    volume = {961},
    pages = {182},
    doi={10.3847/1538-4357/ad118e}
}

@ARTICLE{Frick2010,
       author = {{Frick}, P. and {Stepanov}, R.},
        title = "{Long-term free decay of MHD turbulence}",
      journal = {EPL (Europhysics Letters)},
         year = 2010,
        month = nov,
       volume = {92},
       number = {3},
        pages = {34007},
          doi = {10.1209/0295-5075/92/34007},
       adsurl = {https://ui.adsabs.harvard.edu/abs/2010EL.....9234007F}
}

@article{voronchikhin1985experimental,
  title={Experimental study of the decay of grid turbulence in homogeneous magnetic field},
  author={Voronchikhin, V. and Genin, L. and Levin, V. and Sviridov, V.},
  journal={Magnetohydrodynamics},
  volume={21},
  number={4},
  pages={131--134},
  year={1985},
  doi={}
}

\newpage
\end{document}